\documentclass{article}
\usepackage{graphicx}

\input{tcilatex}
\begin{document}

\textbf{Fractal representation of the power daily demand based on
topological properties of Julia sets.}

H\'{e}ctor A. Tabares-Ospina\footnote{%
Facultad de Ingenier\'{\i}a, Instituci\'{o}n Universitaria Pascual Bravo,
Medell\'{\i}n, Colombia.
\par
Email: h.tabares@pascualbravo.edu.co} and John E. Candelo-Becerra\footnote{%
Facultad de Minas, Universidad Nacional de Colombia, Medell\'{\i}n, Colombia.
\par
Email: jecandelob@unal.edu.co}\vspace{0.5cm}

\bigskip \textbf{Abstract}

In a power system, the load demand considers two components such as the real
power ($P$) because of resistive elements, and the reactive power ($Q$)
because inductive or capacitive elements. This paper presents a graphical
representation of the electric power demand based on the topological
properties of the Julia Sets, with the purpose of observing the different
graphic patterns and relationship with the hourly load consumptions. An
algorithm that iterates complex numbers related to power is used to
represent each fractal diagram of the load demand. The results show some
representative patterns related to each value of the power consumption and
similar behaviour in the fractal diagrams, which allows to understand
consumption behaviours from the different hours of the day. This study
allows to make a relation among the different consumptions of the day to
create relationships that lead to the prediction of different behaviour
patterns of the curves.\vspace{0.5cm}

\textbf{Keywords:} Real power, reactive power, fractal geometry, Julia set,
Mandelbrot set, behavior patterns, power factor

\newpage

\section{Introduction}

Benoit Mandelbrot defined the concept of fractals as a semi-geometric
element with a repetitive structure at different scales [1], with
characteristics of self-similarity as seen in some natural formations such
as snowflakes, ferns, peacock feathers, and Romanesco broccoli. Fractal
theory has been applied to various fields such as biology [2, 3], health
sciences [4, 5, 6, 7, 8], stock markets [9], network communications [10, 11,
12], and others. Fractal theory is one of the methods used to analyse data
and obtain relevant information in highly complex problems. Thus, it has
been used to study the price of highly variable markets, which are not
always explainable from classical economic analysis.

For example, in [9] the authors demonstrate that the current techniques have
some issues to explain the real market operation, and a better understanding
is achieved by using techniques as theory of chaos and fractals. In their
publication, the authors show how to apply the fractal behavior of stock
markets. Besides, the authors refer to multifractal analysis and
multifractal topology. The first describes the invariability of scaling
properties of time series and the second is a function of the H\"{o}lder
exponents that characterize the degree of irregularity of the signal and
their most significant parameters.

In [13], the authors discuss the basic principle of fractal theory and how
to use it to forecast the short-term electricity price. In the first
instance, the authors analyze the fractal characteristic of the electricity
price, confirming that price data have this property. In the second
instance, a fractal model is used to build a forecasting model, which offers
a wide application in determining the price of electricity in the markets.

Similarly, the authors of [14] demonstrate that the price of thermal coal
has multifractal features by using the concepts introduced by
Mandelbrot-Bouchaud. Hence, a quarterly fluctuation index (QFI) for thermal
power coal price is proposed to forecast the coal price caused by market
fluctuation. This study also provides a useful reference to understand the
multifractal fluctuation characteristics in other energy prices.

The fractal geometry analysis has been also applied to study the morphology
and the population growth of cities, and that the electricity demand is
related to the demography of cities. In [15] a multifractal analysis is used
to forecast the electricity demand, explaining that two fractals are found,
showing the behavior pattern of the power demand. Two concepts linked to
fractal geometry are fractal interpolation and extrapolation, related to the
resolution of a fractal-encoded image. In [16], an algorithm to forecast the
electric charge, in which fractal interpolation and extrapolation are also
involved. For the forecast data set, the average relative errors are only
2.303\% and 2.296\%. The result shows that the algorithm has advantages in
improving forecast accuracy.

In the literature reviewed there are no papers focused on the daily real and
reactive power consumption based on the topological properties of the Julia
sets. There are no graphical analyses that show the behaviour of the system
by observing different fractal patterns. Most studies on fractals are
focused on other types of applications such as medicine, biology,
communications, electronics, leading to an important opportunity to perform
the study on power systems. In addition, the characteristics of the real and
reactive powers are not analysed in depth by applying fractal geometry,
concluding that these techniques are not commonly used to study the
different power consumption. Therefore, this paper studies a typical load
demand curve of an electric power system with the fractal theory of Julia
sets. The graphic study focuses on determining the characteristics that the
fractal diagrams created from Julia sets related with the complex numbers of
real and reactive powers and seeking for other graphical patterns of the
load demand curve. For this reason, this work proposes the following
hypothesis: The load demand curve has a clear fractal pattern obtained from
the Julia sets, which allows to characterize the consumption behaviour. The
main contribution of this article is related to the development of a new
methodology, as a complement to those found in literature, which allows to
characterize the daily load demand curve. This paper confirms that the
density of the folds in the Julia sets reveal the transitivity in the
electric power consumption registers, which is related to the irregularity
of the hourly consumption. Eventually, the fractal topology that is obtained
from the Julia sets reflects when the load is inductive or capacitive.

\section{Research method}

\bigskip This methodology is based on constructing an algorithm that allows
an analysis of the fractal diagrams applied to the typical load demand
curve. Below, this section shows the general procedure and the algorithms
implemented to obtain the Mandelbrot and Julia sets, which are useful tools
to graph fractals from the complex numbers related to the load demand.

\subsection{General procedure}

Figure \ref{Fig1} presents a step-by-step procedure applied to graph the
fractal diagrams from the load demand with the Mandelbrot and Julia sets.
This figure shows that the first step ($P1$) is to convert the initial data
to manage the procedure to the Mandelbrot and Julia algorithms. Next, the
Mandelbrot algorithm is programmed according to the mathematical theory ($P2$%
) to generate a new data set. Besides, the Julia algorithm is also
programmed to perform the generation of the new sets, based on the
Mandelbrot set. With these data sets, it is possible to plot the different
fractals ($P4$) which are then analysed to present the different results in
this paper ($P5$) and the corresponding conclusions.

\includegraphics{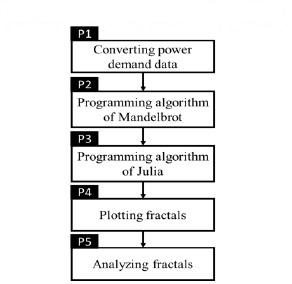}

\bigskip 
\begin{figure*}[h]\centering%
\caption{General procedure of the study}%
\label{Fig1}%
\end{figure*}%

\bigskip \newpage

\subsection{Load demand curve}

The process begins by reading the typical load demand records of real and
reactive defined for a $24$-hour period (see Table 1). The per unit values
of the power demand are calculated with the following expression: $%
per\_unit\_power=actual\_power/base\_power$. In this case, the base power is 
$4000$ MVA. These data are used to plot the diagrams with the programmed
algorithms, in which the lowest and highest consumption points are
considered to evaluate the different fractal diagrams.

\bigskip 
\begin{table*}[ptb]\centering%
\caption{Daily power demand}%
\label{tbl1}%
\begin{tabular}{lllllll}
$Hour$ & $P$ & $Q$ & $S$ & $P_{pu}$ & $Q_{pu}$ & $S_{pu}$ \\ 
00:00:00 & 889 & 371 & 963 & 0.222 & 0.092 & 0.240 \\ 
01:00:00 & 834 & 405 & 927 & 0.208 & 0.101 & 0.231 \\ 
02:00:00 & 792 & 337 & 861 & 0.197 & 0.082 & 0.215 \\ 
03:00:00 & 790 & 324 & 854 & 0.199 & 0.081 & 0.213 \\ 
04:00:00 & 804 & 323 & 867 & 0.201 & 0.080 & 0.216 \\ 
05:00:00 & 925 & 355 & 991 & 0.231 & 0.088 & 0.247 \\ 
06:00:00 & 1041 & 482 & 1147 & 0.260 & 0.120 & 0.286 \\ 
07:00:00 & 1105 & 556 & 1237 & 0.276 & 0.139 & 0.309 \\ 
08:00:00 & 1191 & 610 & 1338 & 0.297 & 0.152 & 0.334 \\ 
09:00:00 & 1256 & 704 & 1439 & 0.314 & 0.176 & 0.359 \\ 
10:00:00 & 1309 & 744 & 1506 & 0.327 & 0.186 & 0.376 \\ 
11:00:00 & 1366 & 775 & 1571 & 0.341 & 0.193 & 0.392 \\ 
12:00:00 & 1385 & 793 & 1595 & 0.346 & 0.198 & 0.398 \\ 
13:00:00 & 1356 & 774 & 1561 & 0.339 & 0.193 & 0.390 \\ 
14:00:00 & 1337 & 759 & 1537 & 0.334 & 0.189 & 0.384 \\ 
15:00:00 & 1350 & 774 & 1556 & 0.337 & 0.193 & 0.389 \\ 
16:00:00 & 1336 & 773 & 1543 & 0.334 & 0.193 & 0.385 \\ 
17:00:00 & 1312 & 749 & 1511 & 0.328 & 0.187 & 0.377 \\ 
18:00:00 & 1287 & 687 & 1459 & 0.321 & 0.171 & 0.364 \\ 
19:00:00 & 1420 & 683 & 1575 & 0.355 & 0.170 & 0.393 \\ 
20:00:00 & 1389 & 660 & 1538 & 0.351 & 0.167 & 0.384 \\ 
21:00:00 & 1311 & 605 & 1444 & 0.327 & 0.151 & 0.361 \\ 
22:00:00 & 1175 & 544 & 1295 & 0.293 & 0.136 & 0.323 \\ 
23:00:00 & 1030 & 489 & 1140 & 0.257 & 0.122 & 0.285%
\end{tabular}%
\end{table*}%

\newpage

\subsection{\protect\bigskip Algorithm to create the Mandelbrot set}

Mandelbrot set, denoted as $M=\{$\ $c\in C/J_{c}\}$, represents the sets of
complex numbers $C$ obtained after iterating the from the initial point $%
Z_{n}$ and the selected constant $C$ as shown (\ref{Ec0}), the results form
a diagram with connected points remaining bounded in an absolute value. One
property of $M$ is that the points are connected, although in some zones of
the diagram it seems that the set is fragmented. The iteration of the
function generates a set of numbers called orbits. The results of the
iteration of those points out of the boundary set tend to infinity.

\begin{equation}
Z_{n+1}=F\left( Z_{n}\right) =Z_{n}^{2}+C  \label{Ec0}
\end{equation}

From the term $C$, a successive recursion is performed with $Z_{0}=0$ as the
initial term. If this successive recursion is dimensioned, then the term $C$
belongs to the Mandelbrot set; if not, then they are excluded. Therefore,
Figure \ref{Fig2} shows the Mandelbrot set with points in the black zone
that are called the prisoners, while the points in other colors are the
escapists and they represent the velocity that they escape to infinite.

\includegraphics{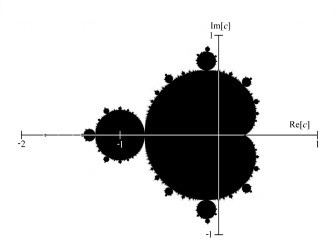}

\bigskip 
\begin{figure*}[h]\centering%
\caption{Representation of Mandelbrot diagram}%
\label{Fig2}%
\begin{tabular}{l}
\end{tabular}%
\end{figure*}%

\bigskip From this figure, the number -1 is inside of the set while the
number 1 is outside of the set. In the Mandelbrot set, the fractal is the
border and the dimension of Hausdorff is unknown. If the image is enlarged
near the edge of the set, many areas the Mandelbrot set are represented in
the same form. Besides, different types of Julia sets are distributed in
different regions of the Mandelbrot set. Whether a complex number appears
with a greater value than 2 in the 0 orbit, then the orbit tends to infinity.

The pseudocode of the algorithm that is used to represent the Mandelbrot set
is presented as follows:

\textbf{Start}

\hspace{1cm}For each point $C$ in the complex plane do:

\hspace{1cm}Fix $Z_{0}=0$

\hspace{1cm}\textbf{For} $t=1$ to $t_{\max }$ \textbf{do}:

\hspace{1cm}\hspace{1cm}Calculate $Z_{t}=Z_{t}^{2}+C$

\hspace{1cm}\hspace{1cm}\textbf{If} $|Z_{t}|>2$ \textbf{then}

\hspace{1cm}\hspace{1cm}\hspace{1cm}Break

\hspace{1cm}\hspace{1cm}\textbf{End if}

\hspace{1cm}\hspace{1cm}\textbf{If} $t<t_{\max }$ \textbf{then}

\hspace{1cm}\hspace{1cm}\hspace{1cm}Draw $C$ in white (the point does not
belong to the set)

\hspace{1cm}\hspace{1cm}\textbf{Else if} $t=t_{\max }$ \textbf{then}

\hspace{1cm}\hspace{1cm}\hspace{1cm}Draw $C$ in black (as the point does
belong to the set)

\hspace{1cm}\hspace{1cm}\textbf{End if}

\hspace{1cm}\textbf{End For}

\textbf{End\vspace{0.5cm}}

In this research, the presented algorithm has been used to obtain the
Mandelbrot set and the diagram that represent it. Some points related to the
real and reactive powers with the respective signs are studied in the
Mandelbrot set and related to those points created for the Julia sets as
explained in the following sections.

\bigskip \newpage

\subsection{Algorithm to create the Julia sets}

At the beginning of the century $XX$, mathematics Gast\'{o}n Julia and
Pierre Fatuo, developed a fractal sets that are obtained by iterating
complex numbers. The Julia sets of a holomorphic function $f$ is constituted
by those points that under the iteration of $f$ have a chaotic behavior and
each point of the set forms a different set $f$ that is then denoted by $%
J(f) $. The Fatou set consists of the points that have a stable behavior
when they are iterated. The Fatou set of a holomorphic function $f$ is
denoted by $F(f)$ and it is a complement of $J(f)$. An important family of
the Julia sets is obtained from the simple quadratic functions, for example $%
Z_{n+1}=F\left( Z_{n}\right) =Z_{n}^{2}+C$, where $C$ is a complex number.
The values obtained from this function are denoted the $J_{c}$, with points
of $Z$ obtained from the parameter $C$ that belong to the Julia sets. Other
points obtained during the iteration are excluded from the Julia sets as
they tend to infinite.

\bigskip For example, in Figure \ref{Fig3}, the complex number $C=0.30+0.21i$
lies within the $M$ set and produces Julia sets of connected points
represented in black, and the points that go to infinity are represented in
different colors according to the number of iterations necessary to escape.

\includegraphics{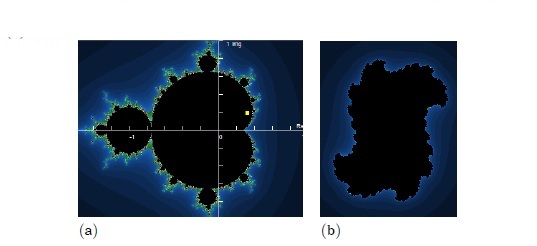}

\begin{figure*}[h]\centering%
\caption{Fractal diagrams when C is inside the M set, plotted as (a) M and (b) J sets}%
\label{Fig3}%
\begin{tabular}{ll}
&  \\ 
(a) & (b)%
\end{tabular}%
\end{figure*}%

\newpage 

However, as shown in Figure \ref{Fig4}, the complex number $C=0.40+0.15i$
lies on the boundary of the $M$ set and produces a Julia set of points that
are partially connected and distributed in different subgroups.

\includegraphics{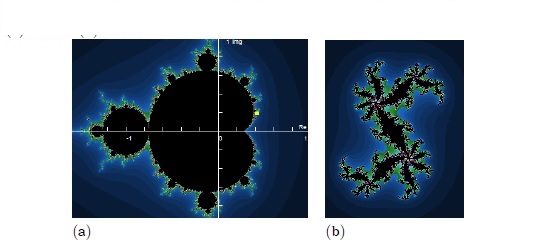}

\begin{figure*}[h]\centering%
\caption{Fractal diagrams when C is in the boundary of the M set, plotted as (a) M and (b) J sets}%
\label{Fig4}%
\begin{tabular}{ll}
&  \\ 
(a) & (b)%
\end{tabular}%
\end{figure*}%

\newpage 

Finally, Figure \ref{Fig5} shows a complete Julia sets of the complex number 
$C=0.50+0.21i$, located outside of the $M$ set.

\includegraphics{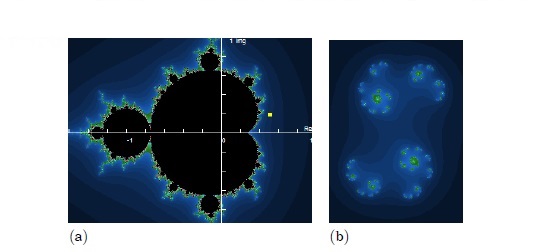}

\begin{figure*}[h]\centering%
\caption{Fractal diagrams when C is outside of M set, plotted as (a) M and (b) J sets}%
\label{Fig5}%
\begin{tabular}{ll}
&  \\ 
(a) & (b)%
\end{tabular}%
\end{figure*}%

\bigskip In summary, there are three types of Julia sets: the first set is
formed by connected points that are obtained when the complex number $C$ is
inside of the Mandelbrot set; the second set is formed by partially
connected points that are obtained when the complex number $C$ is the
boundary of the Mandelbrot set; and the third set is formed by non-connected
points when the constant $C$ is outside of the Mandelbrot set, resulting in
infinite collections of isolated points with no discernible pattern. An
important relation between the Mandelbrot and Julia sets is given when point 
$C$ belongs to the Mandelbrot set; then, the Julia set $J(f_{C})$ obtain a
series of points that are connected. On the other side, when the point does
not belong to the Mandelbrot set, the Julia set $J(f_{C})$ is formed by
non-connected points.

One property of the Mandelbrot set is that the different types of Julia sets
are distributed in different regions of the set $M$. For all the above, it
is concluded that in Figure \ref{Fig5}, $C1$ is in the set of M and $C2$ is
not in the set. In general, for any point within the $M$ cardioid or its
boundary, the Julia set of $J(f_{C})$ has points that are connected. The
most interesting $C$ values are those near the boundary of the Mandelbrot
set because the points can be transformed from connected points to
non-connected points.

\bigskip

\newpage

\subsection{Algorithm to study the fractals of power demand}

In order to obtain the results of the fractal topology patterns that
represent the real and reactive power demand curves, the procedure shown in
Figure \ref{Fig6} was followed.

\includegraphics{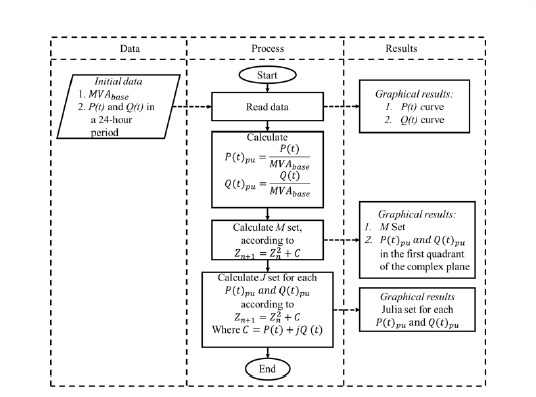}

\begin{figure*}[h]\centering%
\caption{Algorithm with the steps used to obtain the fractal of the power demand}%
\label{Fig6}%
\begin{tabular}{l}
\end{tabular}%
\end{figure*}%

\bigskip The initial process stars reading data of the power base and the
real and reactive powers, followed by the calculation of each per unit
value. In this case, the $P$ and $Q$ curves with respect to the time are
plotted to represent the power demand during the day. Next, the $M$ set is
calculated and used to plot the fractal diagram. Now, the real and reactive
powers of each point in the load demand curve are scaled into the $M$ set
and used to obtain the $J$ sets. Then, the $J$ sets are plotted into fractal
diagrams to analyse qualitatively their geometries.

\newpage 

\section{Results and analysis}

Figure \ref{Fig7} presents the typical demand curves plotted with the data
of Table \ref{tbl1} and Figure \ref{Fig8} presents the power demand plotted
in the first quadrant of the complex plane. As real and reactive powers are
positive, they represent a load consumption related to inductive elements.
Under these conditions, the three most interesting values of the power
consumption are selected such as the lowest consumption at 3:00, the highest
consumption at 19:00, and the approximate average consumption at 09:00.
Other hours of the day represent diagrams that are forms between the values
as shown in the following results in this section.

\includegraphics{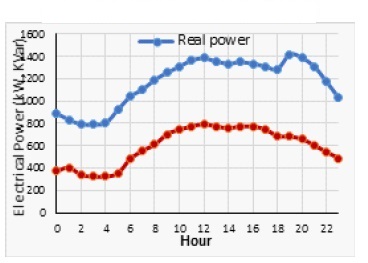}

\begin{figure*}[h]\centering%
\caption{Typical load demand in a day}%
\label{Fig7}%
\begin{tabular}{l}
\end{tabular}%
\end{figure*}%

\newpage 

\includegraphics{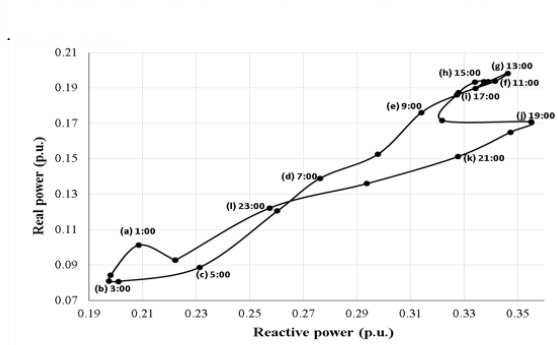}

\begin{figure*}[h]\centering%
\caption{Real and reactive power plotted in the first quadrant of the complex plane of the M set}%
\label{Fig8}%
\begin{tabular}{l}
\end{tabular}%
\end{figure*}%

\newpage 

Figure \ref{Fig9} shows the fractal generated for each point of Figure 8.
These fractals are created by performing iterations of the complex numbers
obtained from the daily load demand.

\includegraphics{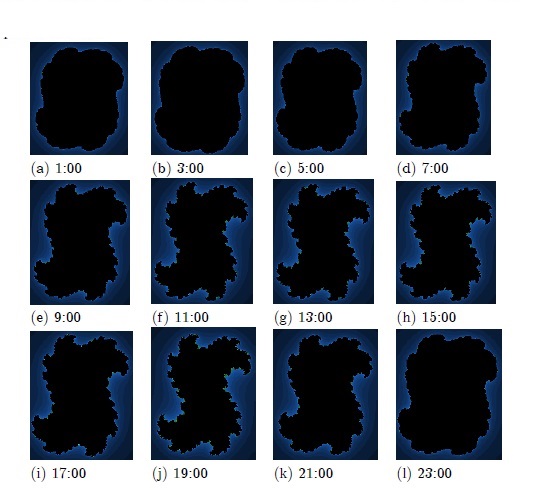}

\begin{figure*}[h]\centering%
\caption{Representation of power demand in the first quadrant of the complex plane of J sets}%
\label{Fig9}%
\begin{tabular}{llll}
&  &  &  \\ 
(a) 1:00 & (b) 3:00 & (c) 5:00 & (d) 7:00 \\ 
&  &  &  \\ 
(e) 9:00 & (f) 11:00 & (g) 13:00 & (h) 15:00 \\ 
&  &  &  \\ 
(i) 17:00 & (j) 19:00 & (k) 21:00 & (l) 23:00%
\end{tabular}%
\end{figure*}%

All Julia sets plotted closed curves of connected points and represent the
fractal topology of inductive power loads that belong to the first quadrant
of the complex plane. The curves are transformed into fractal curves, where
the semi plane save inverted reflections each other and with quadrants
symmetrically inverted with respect to the origin. It is also true that,
when the set of Julia is connected, the point $C$ does not reach the
boundary of the M set, and do not generate periodicity of the Julia set.

Now, with respect to the topological properties of the Julia sets registered
in the lower demand period $[00:00-05:00]$, the equivalent fractal topology
is presented at $3:00$ with an inverted reflective symmetry. Another feature
of the sub-period is the practical invariance in the load demand, which is
evident in the unmoving folding at the boundary of the fractal curve. At $%
05:00$, the load demand begins to increase, generating greater fractal folds
at the boundaries of the Julia set. At $09:00$, the significant increase in
the load demand is evidenced by the increase in the fractal folds of the
Julia set. From $12:00$ to $17:00$, an average load demand is maintained,
representing insignificant changes in the boundary of the Julia set. The
point of greatest interest occurs at $19:00$, when the greatest load demand,
generating a topology with dense fractal folds at the boundary related to
the properties of the peak load demand. From $19:00$ to $24:00$, the load
demand decreases and the boundary of Julia set softens.

\section{Conclusions}

The paper presented a graphical representation of the power demand based on
the topological properties of the Julia Sets, with the purpose of observing
the different graphic patterns and relationship with each consumption in a
daily load demand curve. An algorithm that iterates complex numbers of real
and reactive powers is used to represent each fractal diagram of the
consumption.

It is concluded that the load demand curve presents a clear fractal topology
pattern of the Julia set and the following observations were obtained:

\begin{itemize}
\item A new way of visualizing the state of the power demand curves is
performed by using the fractal diagrams of Julia set.

\item The fractal topology of the Julia sets related to the properties of
power demand does not give a quantitative but qualitative geometry
information as results of studying the different images.

\item The topology of the Julia set reveals that the real and reactive
powers studied for the load demand curve, which is related to an inductive
load, belongs to the first quadrant of the complex plane.

\item The density of folds obtained from the Julia set is related to the
proximity of the demand for real and reactive powers to the boundary of the
Mandelbrot set.

\item From the electrical point of view, the densities of folds are related
to: (a) power consumption through the different hours and (b) the
combination of the real and reactive power magnitudes during the day.
\end{itemize}

The load demand curves studied in this article produce Julia sets with
connected points because the points are within the Mandelbrot set. The
fractals found with the Julia sets evidence that the load demand curves
relate to the steady-state operation, which represents the zone of
predictable values. After repeating the simulation for different real and
reactive powers within the first quadrant of the complex plane, they produce
the Julia set with symmetrically inverted fractal curves with respect to the
origin.

\section{Acknowledgement}

This work was supported by the Agencia de Educaci\'{o}n Superior de Medell%
\'{\i}n (Sapiencia), under the specific agreement celebrated with the
Instituci\'{o}n Universitaria Pascual Bravo. The project is part of the
Energy System Doctorate Program and the Department of Electrical Energy and
Automation of the Universidad Nacional de Colombia, Sede Medell\'{\i}n,
Facultad de Minas.

\section{References}

\hspace{0.5cm}

[1]\qquad Barnsley, M. F., Devaney, R. L., Mandelbrot, B. B., Peitgen, H.
O., Saupe, D., Voss, R. F.. "The Science of Fractal Images". Peitgen, H. O.
\& D. Saupe, Eds. Springer New York, New York, NY, 1998

[2]\qquad Losa, G. A., "Fractals and Their Contribution to Biology and
Medicine". Medicographia, Volume 34, pp. 365--374, 2012.

[3]\qquad Strogatz, S. H., "Nonlinear Dynamics and Chaos". CRC Press, 2018

[4]\qquad Moon, P., Muday, J., Raynor, S., Schirillo, J., Boydston, C.,
Fairbanks, M. S., Taylor, R. P., "Fractal images induce fractal pupil
dilations and constrictions". International Journal of Psychophysiology,
Volume 93, pp. 316--321, 2014.

[5]\qquad Rodr\'{\i}guez V, J. O., Prieto B, S. E., Correa H, S. C.,
Soracipa M, M. Y., Mendez P, L. R., Bernal C, H. J., Hoyos O, N. C., Valero,
L. P., Velasco R, A., Bermudez, E.. "Nueva metodolog\'{\i}a de evaluaci\'{o}%
n del Holter basada en los sistemas din\'{a}micos y la geometr\'{\i}a
fractal: confirmaci\'{o}n de su aplicabilidad a nivel cl\'{\i}nico". Revista
de la Universidad Industrial de Santander - Salud, Volume 48, pp. 27--36,
2016.

[6]\qquad Garcia, T. A., Tamura Ozaki, G. A., Castoldi, R. C., Koike, T. E.,
Trindade Camargo, R. C., Silva Camargo Filho, J. C. "Fractal dimension in
the evaluation of different treatments of muscular injury in rats". Tissue
and Cell, Volume 54, pp. 120--126, 2018.

[7]\qquad Hern\'{a}ndez Vel\'{a}zquez, J. de D., Mej\'{\i}a-Rosales, S.,
Gama Goicochea, A. "Fractal properties of biophysical models of pericellular
brushes can be used to differentiate between cancerous and normal cervical
epithelial cells". Colloids and Surfaces B: Biointerfaces, Volume 170, pp.
572--577, 2018.

[8]\qquad Popovic, N., Radunovic, M., Badnjar, J., Popovic, T. "Fractal
dimension and lacunarity analysis of retinal microvascular morphology in
hypertension and diabetes". Microvascular Research, Volume 118, pp. 36--43,
2018.

[9]\qquad Mandelbrot, B. B., Hudson, R. L. "The (mis) Behaviour of Markets:
A Fractal View of Risk, Ruin and Reward". Profile Books, London, 2004.

[10]\qquad Kumar, R., Chaubey, P. N. "On the design of tree-type ultra
wideband fractal Antenna for DS-CDMA system". Journal of Microwaves,
Optoelectronics and Electromagnetic Applications, Volume 11, pp. 107--121,
2012.

[11]\qquad Ye, D., Dai, M., Sun, Y., Su, W., "Average weighted receiving
time on the non-homogeneous double-weighted fractal networks". Physica A:
Statistical Mechanics and its Applications, Volume 473, pp. 390--402, 2017.

[12]\qquad Ma, Y. J., Zhai, M. Y. "Fractal and multi-fractal features of the
broadband power line communication signals". Computers \& Electrical
Engineering. In Press, Corrected Proof, 2018.

[13]\qquad Cui, H., Yang, L. "Short-Term Electricity Price Forecast Based on
Improved Fractal Theory". In: IEEE International Conference on Computer
Engineering and Technology, pp. 347--351, 2009.

[14]\qquad Zhao, Z. yu, Zhu, J., Xia, B. "Multi-fractal fluctuation features
of thermal power coal price in China". Energy, Volume 117, pp. 10--18, 2016

[15]\qquad Salv\'{o}, G., Piacquadio, M. N. \textquotedblleft Multifractal
analysis of electricity demand as a tool for spatial
forecasting\textquotedblright . Energy for Sustainable Development, Volume
38, pp. 67--76, 2017.

[16]\qquad Zhai, M. Y. \textquotedblleft A new method for short-term load
forecasting based on fractal interpretation and wavelet
analysis\textquotedblright . International Journal of Electrical Power \&
Energy Systems, Volume 69, pp. 241--245, 2015.

\section{Bibliography of authors}

\includegraphics{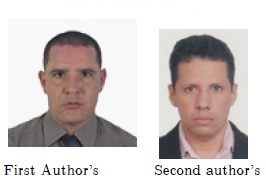}

\begin{figure*}[h]\centering%
\caption{Authors}%
\label{Fig10}%
\begin{tabular}{ll}
&  \\ 
First Author's & Second author's%
\end{tabular}%
\end{figure*}%

Firts Author's: H\'{e}ctor A. Tabares-Ospina: received his Bs. degree in
Electrical Engineering in 1997 and his Master in Systems Engineering in 2005
from Universidad Nacional de Colombia. He is now studing doctoral studies.
He is an Assistant Professor of Instituci\'{o}n Universitaria Pascual Bravo.
His research interests include: Fractal geometry, artificial intelligence,
operation and control of power systems; and smart grids. He is a Junior
Researcher in Colciencias and member of the Research Group - GIIEN, at
Instituci\'{o}n Universitaria Pascual Bravo.
https://orcid.org/0000-0003-2841-6262\\*[0pt]

Second Author's:John E. Candelo-Becerra: received his Bs. degree in
Electrical Engineering in 2002 and his PhD in Engineering with emphasis in
Electrical Engineering in 2009 from Universidad del Valle, Cali - Colombia.
His employment experiences include the Empresa de Energ\'{\i}a del Pac\'{\i}%
fico EPSA, Universidad del Norte, and Universidad Nacional de Colombia -
Sede Medell\'{\i}n. He is now an Assistant Professor of the Universidad
Nacional de Colombia - Sede Medell\'{\i}n, Colombia. His research interests
include: engineering education; planning, operation and control of power
systems; artificial intelligence; and smart grids. He is a Senior Researcher
in Colciencias and member of the Applied Technologies Research Group - GITA,
at the Universidad Nacional de Colombia.
https://orcid.org/0000-0002-9784-9494.

\end{document}